\documentclass{pasj00}
\draft

\newcommand{\ergcs}{erg~cm$^{-2}$s$^{-1}$}
\newcommand{\ergs}{erg~s$^{-1}$}
\def\ltsima{$\; \buildrel < \over \sim \;$}
\def\simlt{\lower.5ex\hbox{\ltsima}}
\def\gtsima{$\; \buildrel > \over \sim \;$}
\def\simgt{\lower.5ex\hbox{\gtsima}}

\begin{document}
\SetRunningHead{Y. Tanaka et al.}
{Partial Covering Interpretation of the NLS1 1H~0707--495}
\Received{2004 March 10}
\Accepted{2004 April 9}

\title{Partial Covering Interpretation of the X-Ray Spectrum of the NLS1 
       1H~0707--495}

\author{Yasuo {\sc Tanaka}, Thomas {\sc Boller}, Luigi {\sc Gallo} 
\& Ralph {\sc Keil}\\
{\it Max--Planck--Institut f\"ur Extraterrestrische Physik,
 Giessenbachstra{\ss}e, 85748 Garching, Germany}\\
 and\\
Yoshihiro {\sc Ueda}\\
{\it Institute of Space and Astronautical Science, 3-1-1 Yoshinodai, 
Sagamihara, Kanagawa 229, Japan}
}



\maketitle

\begin{abstract}
The X-ray spectrum of 1H~0707--495 obtained with XMM-Newton showing 
a deep flux drop at $\sim7$ keV (Boller et al. 2002) is 
studied based on the partial covering concept. The previously inferred extreme 
iron overabundance can be reduced down to $\sim3\times$ solar if the hard 
component gradually steepens at high energies. The spectral shape 
supports that 1H~0707--495 is an AGN analogue of the galactic black-hole 
binaries in the soft state. Interpreting the soft excess as the emission from 
an optically-thick disk, the minimum black hole mass $M$ is estimated to be 
$2\cdot10^6$ $M_{\odot}$ from the intrinsic luminosity corrected 
for partial covering. Based on the slim disk model, the observed disk 
temperature implies that the luminosity is close to 
the Eddington limit. The rapid and large flux variations with little change 
in the spectral shape can also be explained, if not all, as due to changes 
in the partial covering fraction. Partial covering may account for the large 
variability characteristics of NLS1. 
\end{abstract}

\section{Introduction}

The X-ray spectrum of the narrow-line Seyfert 1 galaxy (NLS1) 1H~0707--495 
($z=0.0411$) observed with XMM-Newton revealed a pronounced flux 
drop by a factor of $\simgt3$ at $\sim7$ keV (Boller et al. 2002, hereafter 
referred to as Paper 1).
The measured edge energy and its sharpness suggest K-absorption by essentially 
neutral iron. However, there is no detectable fluorescence line at 6.4 keV 
(rest frame). 
In contrast to the strong iron K-absorption, the soft excess  
dominating below $\sim 1$ keV is little absorbed. 

Two possibilities for these distinct features were discussed 
in Paper 1, i.e. (1) partial covering and (2) reflection from an ionized disk. 
Both models require an extreme iron overabundance for the large depth of the 
edge ($\sim35\times$ solar for partial covering of a power law model). 
The disk reflection model has been elaborated on by Fabian 
et al. (2002), considering reflections on sheets of dense material presumably 
formed by disk instabilities.  
Their model can account for the observed spectral features 
with an iron abundance of $\sim5\times$ solar. 

In this paper, we pursue the partial covering concept. The observed 
spectral features (little low-energy absorption, deep iron K-edge, and 
the absence of a prominent fluorescence line) are consistent with a partial 
covering phenomenon. In fact, similar features were observed in  
galactic X-ray binaries, and explained as due to partial 
covering (Brandt et al. 1996; Tanaka et al. 2003). 

The spectral shape of 1H~0707--495 (an intense soft component and a
hard tail) is similar to those of the galactic black-hole 
binaries in the soft state at high-luminosities (Tanaka, Shibazaki 1996), 
except that the soft excess lies in a much lower energy range. 
We consider that the soft excess of 1H~0707--495 is the emission from an 
optically-thick accretion disk as in the case of the soft-state black-hole 
binaries, and analyze the spectrum based on the partial covering concept 
with particular attention to the iron overabundance problem. 

It should be mentioned that the model described in this paper has also been 
applied to the analysis of the XMM-Newton data of another NLS1 
IRAS~13224--3809 (Boller et al. 2003) which showed a similar spectral shape and 
a deep flux drop as observed in 1H~0707--495. 

Using the results of the analysis, we try to constrain the central black hole 
mass of 1H~0707--495. 
A possibility that the fast X-ray variability is due to changing 
covering fraction is also discussed.


\section{Spectral Analysis}

The XMM-Newton observation of 1H~0707--495 and the data processing are 
described in Paper 1. The X-ray spectrum shown in Paper 1 is used for 
the present analysis. 

The observed spectrum of the soft excess is well reproduced by the 
multi-color blackbody disk (MCD) model (Mitsuda et al. 1984; Makishima et al. 
2000) that approximates the spectrum of an optically-thick disk. 
A single-temperature blackbody model also gives an equally good fit, which is 
expected since the downward slope of MCD has a blackbody shape. 

\subsection{Low-Energy Absorption}

First, we determine the low-energy absorption from the data in the 0.3--2 keV 
range. Simultaneous fitting is performed on all of the  
EPIC (MOS1, MOS2, and pn) spectra, utilizing a model 
consisting of a MCD or a blackbody and a power-law, modified by absorption. 
The parameter values for all three spectra are kept the same except for the 
normalization. 

Simple absorption by neutral gas of cosmic abundances is found to be inadequate. 
An acceptable fit is obtained 
if an absorption edge is added at $\sim 0.37$ keV. This may be the K-edge of 
C~V (0.39 keV, rest frame) of a warm absorber. 
For a MCD model, we obtain an absorption column of $\sim 1.3\times10^{21}$ 
cm$^{-2}$ including the interstellar absorption, and the depth of the 0.37-keV 
edge $\tau \sim 1.3$. A blackbody model gives slightly less 
absorption, $\sim 1.1\times10^{21}$ cm$^{-2}$ and $\tau \sim 1.2$. 
Note that these values are still subject to calibration uncertainties in the 
low-energy region. 
The best-fit MCD temperature (the highest color temperature of the disk) is 
$kT_{\rm in} \sim 100$ eV, and $kT_{\rm bb}\sim$92 eV for 
the blackbody temperature. 
The power-law photon index $\Gamma$ is $\sim2.4$ for both models. 
For the later analysis we fix these absorption parameters. 

\subsection{Partial Covering Spectral Analysis} 

We analyze the spectrum over the range 0.6--12 keV. We use only 
the pn data for its high detection efficiency at high energies.
Exclusion of the data below 0.6 keV does not influence the fitting of the 
soft excess when the absorption parameters are fixed as determined above. 

We employ the partial covering model with a neutral absorber as indicated by 
the observed energy of the iron K-edge. 
When a single absorber is responsible for partial covering, the fitting model 
is expressed as\\

 $[f e^{-\sigma N_{\rm H}} + (1-f)][F_{\rm soft}(E) + F_{\rm hard}(E)]$ , 
 \hspace{1cm} (1)\\
 

\noindent
where $f$ denotes the covering fraction by an absorber of a column density 
$N_{\rm H}$, and $\sigma$ the photoabsorption cross section containing the Fe  
abundance as a free parameter. 
$F_{\rm soft}(E)$ and $F_{\rm hard}(E)$ are the model 
functions of the soft and hard components, respectively. 

\begin{figure}[t]
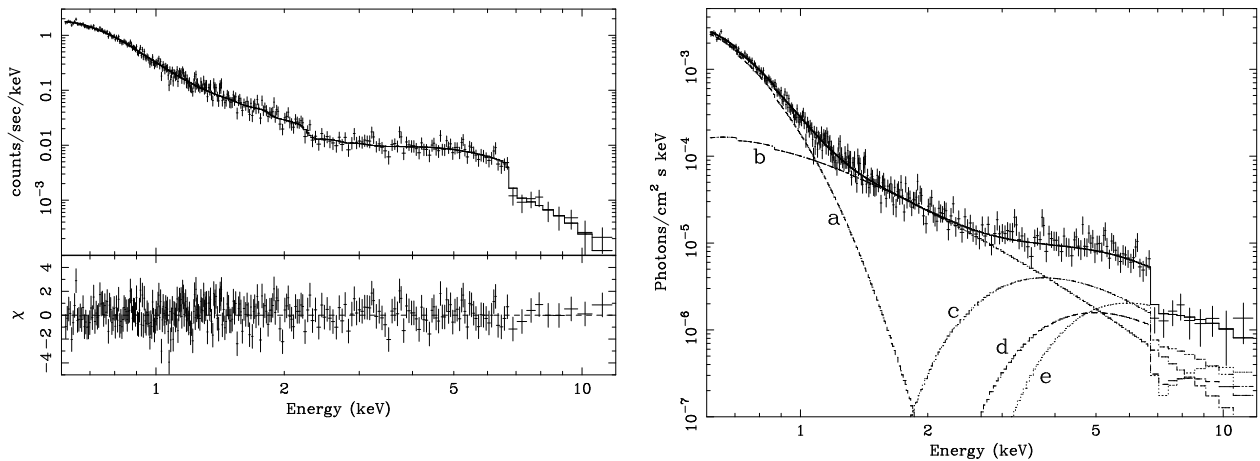

 \begin{center}
   \rotatebox{270}{\FigureFile(54.5mm,){figure1a.eps}}
   \rotatebox{270}{\FigureFile(60mm,){figure1b.eps}}
 \end{center}
\caption{Double partial covering applied to the model consisting of a MCD 
and a cut-off power law. The left panel is the best fit (Fe abundance 
$13\times$ solar) to the observed spectrum, and the right panel is the 
unfolded spectrum showing the model components: (a) the MCD and (b) the 
cut-off power law. (c), (d) and (e) are those absorbed with column densities 
$N_{\rm H}$, $N_{\rm H,2}$, and $N_{\rm H}+N_{\rm H,2}$, respectively (see 
Formula~(2)).}
\end{figure}

If a straight power law is employed for $F_{\rm hard}(E)$, an extreme Fe 
abundance ($\sim50\times$ solar) is required. The situation improves 
substantially if double partial covering by two separate absorbers is 
considered and expressed as\\

 $[f e^{-\sigma N_{\rm H}} + (1-f)][f_2 e^{-\sigma N_{\rm H,2}} + (1-f_2)]$ 

 $\times [F_{\rm soft}(E) + F_{\rm hard}(E)]$ , 
 \hspace{3.6cm} (2)\\
 

\noindent
where $f_2$ is the covering fraction of the second absorber of a column 
density $N_{\rm H,2}$. Double partial covering is also an 
approximation for the sum of two absorbed components with different column 
densities, as well as for more complex cases involving multiple absorbed 
components. 
Evidence for double partial covering was found in the spectrum of the X-ray 
binary GRO~1655--40 during a dip-like event (Tanaka et al. 2003). 

A double partial covering gives a good fit for a power-law model in 
1H~0707--495 with $\chi ^2_{\rm min}/\nu=301/293$. 
The best-fit Fe abundance is as large as $30\times$ solar, 
similar to the result of Paper I. The acceptable range 
within 90\% limit ($\chi ^2_{\rm min} +2.71$) extends down to $5\times$ solar, 
although the power-law becomes very steep with a photon index $\Gamma\sim3.4$ 
(see table~1). 

We notice a general trend that $\Gamma$ increases as the Fe abundance is 
reduced. 
This suggests that gradual steepening of the hard tail toward high energies 
would allow for lower Fe abundances. For such a spectral model, we 
employ a cut-off power law of the form $E^{-\Gamma}e^{-E/E_{\rm c}}$.  

First, when single partial covering [Formula (1)] is assumed, 
we find a shallow minimum around the best-fit Fe abundance of $26\times$ solar 
($\chi ^2_{\rm min}/\nu=301/294$).  
Fe abundance down to $5\times$ solar is acceptable at 90\% limit (see table~1). 
The main difference from the simple power-law case is that the average $\Gamma$ 
below $\sim 5$ keV stays at around $\sim2.5$ over the wide range of Fe 
abundance. 

\begin{table*}
 \centering
 \begin{minipage}{17.0cm}
 \caption{Results of the fitting with the partial covering model.}

\begin{tabular*}{17.0cm}{ll@{\extracolsep\fill}llllllll}
\hline\hline
\noalign{\smallskip}
Hard component& &Fe abundance&$\Gamma$& $E_{\rm c}$& \ \ \ $f$&$ \ N_{\rm H}$& 
$ \ f_2$&  $N_{\rm H,2}$& $\chi^2/\nu$\\

\noalign{\smallskip}
&&$\times$ solar& &keV&& $10^{22}$ cm$^{-2}$ &&$10^{22}$ cm$^{-2}$& \\

\noalign{\smallskip}
\hline
\noalign{\smallskip}
Power law & &30 \ best fit&
      2.7&$\infty$& \ \ 0.39& \ 0.7&0.84& \ 4.6 & 301/293\\

      && \ 5&3.4&$\infty$& \ \ 0.66& \ 1.8&0.90&17.5&304/294\\

\noalign{\smallskip}
\hline
\noalign{\smallskip}
Cut-off power law & &26 \ best fit& 2.2& 9.2& \ \ 0.85& \ 5.1& \ $-$& \ $-$&
      301/294\\

      && \ 5& 1.9& 4.2& \ \ 0.88& 16.7& \ $-$& \ $-$& 304/295\\

\noalign{\smallskip}
\hline
\noalign{\smallskip}
Cut-off power law & &13 \ best fit& 
      2.3&8.0& \ \ 0.62& \ 4.2&0.73&10.3& 297/292\\
      & & \ 5&  2.2&4.5& \ \ 0.71& \ 7.8&0.76&22.7& 298/293\\
      & & \ 2.5&2.0&2.8& \ \ 0.78& 11.4 &0.79&37.9& 300/293\\ 

\noalign{\smallskip}
\hline
\end{tabular*}
\end{minipage}
\end{table*}

Double partial covering [Formula (2)] with a cut-off power law model allows for 
even smaller Fe abundances. While the best-fit value is 
$\sim13\times$ solar ($\chi ^2_{\rm min}/\nu=300/292$), the 90\%-confidence 
lower limit is $2.5\times$ solar (see table~1). The average $\Gamma$ in the 
range 2--5 keV is also $\sim2.5$. 
The best spectral fit is shown in figure 1.

The parameter values obtained from these fits with a MCD soft component 
model are listed in table~1. A blackbody model gives essentially the same 
results. Differences in the hard component model do not affect the fits of  
the soft component because the latter dominates the observed spectrum.   
The absorption-corrected soft component accounts for $\sim$90\% of the total 
flux above 0.3 keV. On the other hand, the intrinsic flux 
(corrected for covering) depends on the covering model (single or double) and 
the Fe abundance, which cause differences in the covering fraction 
(see table~1).

\section{Discussion} 

The observed spectral shape of 1H~0707--495 showing an intense soft  
component and a hard tail is strikingly similar to those of the black-hole 
binaries in the soft state. The soft component which displays a 
blackbody-shaped spectrum and lack of prominent line features is consistent 
with the emission from an optically-thick disk. The hard component shows an 
average photon index $\Gamma \sim2.5$, similar to 
the hard tail of the black-hole binaries in the soft state, and distinctly 
softer than in the hard state ($\Gamma \sim 1.8)$.   

These properties are fairly common among NLS1 (e.g. Boller 2000),  
suggesting that they are AGN analogues of the soft-state black-hole binaries. 
This was first pointed out for the NLS1 RE~1034+39 by Pounds et al. (1995).  
In contrast, most other AGNs show harder power-law spectra, analogous 
to the black-hole binaries in the hard state at low luminosities.

\subsection{Partial Covering and Iron Abundance}

The observed spectral features (little low-energy absorption and deep Fe 
K-absorption without a prominent fluorescence line) are typical of partial 
covering. That is, the central source is covered by absorbing clouds in the 
line of sight; however, either the absorber is patchy so that a fraction of 
the emission leaks through, or the photons which are electron-scattered off 
the source form the unabsorbed component. 

Table 1 shows that the covering fraction is roughly 90\%, i.e. 
the intrinsic source flux is an order of magnitude higher than the observed 
time-averaged flux (corresponding to the spectrum analyzed here). This is 
not surprisingly high, considering that the peak flux during this observation 
was $\sim$3 times the average (see figure~3 of Paper 1).   

In Paper 1 an excessively large ($\sim30\times$ solar) Fe abundance was required 
for partial covering of a power-law model. However, the present 
analysis shows that the Fe abundance depends on the shape of the hard 
continuum, which is still uncertain. 
Gradual steepening of the hard component (modelled with a cut-off power law) 
allows for lower Fe abundances. 
Double partial covering gives the lowest acceptable Fe abundance of 
2.5$\times$ solar, which is still suggestive of substantial overabundance. 
Accurate determination requires much better statistics above 
the edge to constrain the shape of the hard component. 

A similarly deep flux drop was discovered from IRAS~13224--3809, also 
attributable to the Fe-K edge (Boller et al. 2003). 
The same partial covering model as used here also gives an Fe 
overabundance of $\sim10\times$ solar (the 90\% confidence lower limit being 
$2\times$ solar) and requires a large covering fraction of $\simgt80$\% 
(see Boller et al. 2003 for detail). 

Such a deep flux drop (by a factor of $>3$) has rarely been observed. Whether 
or not these two sources are exceptional among NLS1 is an interesting question. 
In this respect, it is to be noticed that the depth of the edge changes 
with the covering fraction, i.e., the larger the covering fraction, the deeper 
the edge becomes (see figure~2 in subsection 3.3). 
Therefore, in addition to the effect of Fe overabundance, the edges of these 
two NLS1 might be deepened due to particularly large covering fraction  
compared to other NLS1 observed. 
Thus, the question still remains open at present. (Note that the covering 
fraction may change with time.)

\subsection{Mass of the Central Object and Accretion Rate}

We try to constrain the mass of the central object in 1H~0707--495 from the 
bolometric luminosity $L_{\rm bol}$ and the innermost disk (color) temperature 
$T_{\rm in}$. 
For estimating $L_{\rm bol}$, the MCD model is valid only for  
$L_{\rm bol}\ll L_{\rm E}$, where $L_{\rm E}$ is the Eddington luminosity. 
As $L_{\rm bol}$ approaches $L_{\rm E}$ (which is the case as shown below), 
the spectral energy distribution deviates from the MCD model due to 
enhanced Comptonization in the innermost disk (e.g. Ross et al. 1992); thus 
the MCD model overestimates $L_{\rm bol}$. We evaluate here the minimum 
value of $L_{\rm bol}$ by employing the present result for a single blackbody 
model instead of a MCD model. 

For the Fe abundance range of $(2.5-5)\times$ solar, the bolometric blackbody 
flux corrected for covering (according to table~1) is in the range 
$(1.0-1.5)\cdot 10^{-10}$ \ergcs (excluding the steep power-law case of 
$\Gamma=3.4$, see table~1). Furthermore, we consider a face-on disk for 
estimating a minimum bolometric luminosity $L_{\rm min}$. 
Thus-obtained $L_{\rm min}$ is $\sim (2-3)\cdot 10^{44}$ \ergs 
for $z=0.0411$ and $H_{\rm 0}=70$~km~s$^{-1}$Mpc$^{-1}$. Assuming 
$L_{\rm min} < L_{\rm E}$, the minimum mass of the central object 
$M_{\rm min}$ is around $2\cdot 10^6$$M_{\odot}$. 

On the other hand, the observed disk temperature (100 eV for a MCD and 92 eV 
for a blackbody) is very high for a standard disk around a Schwarzschild hole 
of such a mass. If the MCD model is used, adopting a color correction factor 
($T_{\rm in}/T_{\rm eff}$) of 2.4 (Ross et al. 1992), the observed temperature 
exceeds the limit ($\sim80$ eV) at $L_{\rm bol}=L_{\rm E}$ [from equation (12) 
of Makishima et al. 2000]. Although the result from the MCD model may no longer 
be accurate near $L_{\rm E}$, it clearly indicates an extremely high 
accretion rate.

A slim disk model, first proposed by Abramowicz et al. (1988), 
applies at such high accretion rates as $\dot M \gg L_{\rm bol}/c^2$. 
In this case, significant radiation can come from within 3 Schwarzschild radii, hence 
$T_{\rm in}$ may exceed the limit of a standard thin disk. Detailed spectral 
properties of a slim disk have been studied 
(e.g. Mineshige et al. 2000; Watarai et al. 2001).

Based on this slim disk model, we extrapolate figure~1 of Watarai et al. (2001), 
which displays the $L_{\rm bol}-T_{\rm in}$ relation for different $M$-values 
of a Schwarzschild hole, and obtain the minimum mass 
$M_{\rm min} \sim 2\cdot10^6$$M_{\odot}$ and $L_{\rm min}\approx L_{\rm E}$  
which corresponds to $\dot M \sim 16(L/c^2)$. 
Considering that true $L_{\rm bol}>L_{\rm min}$, $M$ will be 
a little larger than $2\cdot10^6$$M_{\odot}$, and then $L_{\rm bol}$ 
will slightly exceeds $L_{\rm E}$ for the observed disk temperature.

\subsection{Time Variability}

As discussed above, the X-ray spectrum of 1H~0707--495 is characteristic of 
the soft state. However, it displays extreme variability on time scales as 
short as $10^3$ s. This is distinctly different from the black-hole X-ray 
binaries in the soft state in which the variability is relatively small.  
(Tanaka, Shibazaki 1996).
Importantly, as reported in Paper 1, 1H~0707--495 shows insignificant  
spectral variability during the large flux variations of more 
than a factor of 4. For an optically-thick disk to change luminosity this 
much, the disk temperature $T_{\rm in}$ must change quite significantly, 
$\simgt40$\%, which is not observed. 
The same holds true for another highly-variable NLS1 IRAS~13224--3809 which  
showed even larger flux variations than 1H~0707--495 without changing 
$T_{\rm in}$ (Boller et al. 2003; Gallo et al. 2004). 
We therefore suspect that the observed flux variations are not all due to 
the luminosity changes of the central source.

\begin{figure}
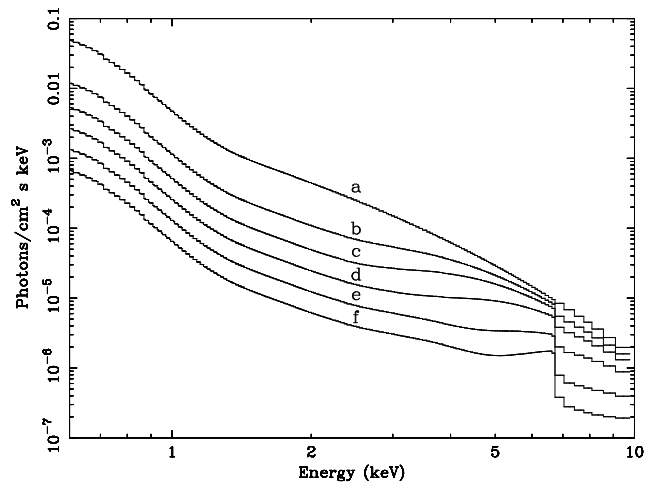

 \begin{center}
  \rotatebox{270}{\FigureFile(63mm,){figure2.eps}}
 \end{center}
\caption{Spectra expected from double partial covering of a MCD + a cut-off 
power law model for different $f_2$ and $N_{\rm H,2}$. 
(a) is the intrinsic spectrum ($f=f_2=0$), and (b)--(f) are for the sets 
of ($f_2$,  
$N_{\rm H,2}$ in units of $10^{22}$ cm$^{-2}$) (0, 0), (0.56, 16), 
(0.78, 33), 
(0.89, 60) and (0.94, 90), respectively.}
\end{figure}

The possibility of partial covering causing large flux changes in NLS1 was 
previously considered (Boller et al. 1997; Brandt \& Gallagher 2000). 
Tanaka et al. (2003) demonstrated, from a 
study of a dip in the light curve of GRO~J1655-40, that spectrum-invariant 
flux changes can occur as a result of changes in partial covering conditions. 
A similar condition is illustrated in figure~2 for the case of 1H~0707--495.  
The parameters of the MCD and of one partial coverer ($f$ and 
$N_{\rm H}$) are fixed to the best-fit values of the observed spectrum 
(Fe abundance of $3\times$ solar assumed), and those of the second 
coverer ($f_2$ and $N_{\rm H,2}$) are varied. Among them, (a) is the intrinsic 
spectrum and (d) is the best fit to the observed spectrum. 
The spectral shapes of (b) through (f) are similar to 
each other over a flux range of an order of magnitude. Positive 
correlation between $f_2$ and $N_{\rm H,2}$ is plausible, as discussed in 
Tanaka et al. (2003). 
  
The observed light curve of 1H~0707--495 (figure~3 of Paper 1) appears as 
consisting of random ``flares" of various peak fluxes, each lasting typically  
$\sim3\cdot10^3$~s. 
Such flux changes can be explained qualitatively as follows:    
If the azimuthal distribution of the covering clouds is non-uniform, 
the covering fraction varies according to the orbital motion 
of the clouds.  
A local minimum of the covering fraction corresponds to a flare peak. 
  
The duration of a flare is the time that a local minimum of covering takes to 
scan the electron-scattering region from a distance $r$ at 
the Keplerian velocity $c(r_{\rm g}/r)^{1/2}$, where  
$r_{\rm g}$ is the gravitational radius.
Suppose the electron-scattering region has an extension of $\sim10r_{\rm g}$,    
then the observed time scale is understood when these clouds are located around 
$r\sim400r_{\rm g}$ ($\sim2\cdot10^{14}$ cm) for $M\sim3\cdot10^6M_{\odot}$. 
 
The essential neutrality of iron indicates low photoionization 
effect, i.e. the ionization parameter $\xi=L_{\rm h}/nr^2 \simlt1$, where 
$L_{\rm h}$ is the hard component luminosity and $n$ the atomic density. 
This requires that $n$$\simgt10^{14}$ cm$^{-3}$ for an estimated 
$L_{\rm h}\sim10^{43}$ erg~s$^{-1}$. 
Such dense clouds are probably confined by magnetic fields (Rees 1987). 
In fact, the results from magneto-hydrodynamical simulations that demonstrate 
highly inhomogeneous and clumpy accretion flow are consistent with this view 
(e.g. Kawaguchi et al. 2000).  

The persistent variation of the observed flux of 1H~0707--495 may be related 
to very high, possibly super-Eddington, luminosity (see subsection 3.2). 
At these accretion rates, various disk instabilities might produce  
inhomogeneities in the vertical direction, causing time-variable partial 
covering. 
An alternative possibility is that partial covering clouds may be formed from 
a radiation-driven mass outflow. 

Large-amplitude variability is generally characteristic of NLS1. 
It is worth examining the associated spectral variation of other NLS1. 
If $T_{\rm in}$ of the soft component does not change with flux, it is 
possible that the flux variation may, at least in part, be due to changes 
in partial covering. 
If it were the case, the intrinsic luminosity, hence 
the black hole mass, could be significantly higher than estimated from 
the observed flux.\\

The authors are grateful to S. Mineshige for valuable comments.

\end{document}